\documentclass[twocolumn,prl]{revtex4}
\usepackage{amssymb}
\usepackage{graphicx}

\input{tcilatex}

\begin{document}

\title{Low temperature fullerene encapsulation in single wall carbon nanotubes: synthesis of N@C$_{60}$@SWCNT} 
\author{$^{1}$F. Simon}
\author{$^{1}$H. Kuzmany}
\author{$^{2}$H. Rauf}
\author{$^{2}$T. Pichler}
\author{$^{3}$J. Bernardi}
\author{$^{1}$H. Peterlik}
\author{$^{4}$L. Korecz}
\author{$^{4}$F. F\"{u}l\"{o}p}
\author{$^{4}$A. J\'{a}nossy}

\address{$^{1}$ Institut f\"{u}r Materialphysik, Universit\"{a}t Wien, Strudlhofgasse 4, A-1090 Wien, Austria}
\address{$^{2}$ Institut f\"{u}r Festk\"{o}rper- und Werkstoffforschung, P.O.Box 270016, D-01171, Dresden, Germany}
\address{$^{3}$ University Service Centre for Transmission Electron Microscopy (USTEM), Technische Universit\"{a}t Wien, Wiedner Hauptstrasse 8 - 10 / 052, A-1040 Wien,  Austria}
\address{$^{4}$ Budapest University of Technology and Economics, Institute of Physics and Solids in Magnetic Fields Research Group of the Hungarian Academy of Sciences, H-1521, Budapest P.O.Box 91, Hungary}

\begin{abstract}
High filling of single wall carbon nanotubes (SWCNT) with C$_{60}$ and C$_{70}$ fullerenes in solvent is reported at 
temperatures as low as 69 $^{o}$C. A 2 hour long refluxing in n-hexane of the mixture of the fullerene and 
SWCNT results in a high yield of C$_{60}$,C$_{70}$@SWCNT, fullerene peapod, material. The peapod filling is characterized 
by TEM, Raman and electron energy loss spectroscopy and X-ray scattering. We applied the method to synthesize 
the temperature sensitive (N@C$_{60}$:C$_{60}$)@SWCNT as proved by electron spin resonance spectroscopy. The solvent 
prepared peapod samples can be transformed to double walled nanotubes enabling a high yield and industrially 
scalable production of DWCNT.
\end{abstract}


\maketitle

\section{Introduction}
Nanostructures based on carbon nanotubes \cite{IijimaNAT} have been in the forefront of nanomaterial research 
in the last decade. Single wall carbon nanotube (SWCNT) is an even more exciting material as it represents the 
perfect, one-dimensional form of carbon. Fullerene encapsulating SWCNTs have attracted considerable interest 
after the discovery of C$_{60}$@SWCNT peapods \cite{SmithNAT}. More recently, several molecules have been 
successfully inserted into the interior of tubes such as other fullerenes, endohedral metallofullerenes or 
alkali halides\cite{MonthiouxCAR}. It is believed that the inside filled structures can alter or 
enhance the mechanical and electronic properties of the SWCNTs or may allow the fine tuning of these parameters.
However, all these synthesis methods required treatment at relatively high temperatures, above 400 $^{o}$C. 
In particular, the peapod synthesis requires the heat treatment of SWCNT and fullerenes sealed together under 
vacuum, a method that appears impractical for large scale production purposes. Another important trend is the 
study of the behavior of the encapsulated materials under special conditions. It was recently shown that 
fullerene peapods are transformed into a double wall carbon nanotube (DWCNT) structure after high temperature 
annealing\cite{BandowCPL}. The fullerenes coalesce into an inner nanotube, which leaves the electronic 
properties unaffected but is expected to significantly enhance the mechanical properties of the tube system. 
This enhanced mechanical stability makes DWCNTs promising candidates for applications such as future electronics, 
probe tips for scanning probe microscopy, field emission devices and many more. 
Our aim in the current study was two-fold: i.) development of a peapod synthesis method that allows the use of 
low temperatures in order to obtain encapsulated materials which do not survive the usual high temperature 
synthesis methods, ii.) devising a simple method for the production of peapod starting materials to facilitate 
the large scale synthesis of DWCNT.
In what follows, we describe the synthesis of fullerene peapods from SWCNT mixed with fullerene in solution. 
We present transmission electron microscopy, Raman spectroscopy, electron energy loss spectroscopy, and 
X-ray studies to prove that a high filling content of the peapods is achieved. We show electron spin resonance 
evidence that the 
temperature sensitive N@C$_{60}$ survives the filling. We also present the transformation of the solvent 
prepared peapod samples into DWCNT.

\section{Experimental}
\textit{Sample preparation.} Commercial SWCNT (NCL-SWCNT from Nanocarblab, Moscow, Russia\cite{nanocarblab} and Rice-SWCNT 
from Tubes@Rice, Rice University, Houston, Texas) and  fullerenes (Hoechst AG, Frankfurt, 
Germany), and n-hexane (Merck KGaA, Darmstadt, Germany) were used for the low temperature synthesis of fullerene 
peapods. The NCL-SWCNT material is prepared by the arc-discharge method and is purified to 50 \% using repeated 
high temperature air and acid washing treatments by the manufacturer. The Rice SWCNT material had an initial 
purity of 15 \% and is purified with a triple repetition of H$_{2}$O$_{2}$ refluxing and HCl acid etching. 
The material was then filtered and degassed in dynamic vacuum at 400 $^{o}$C for 1 h. The filling levels 
discussed below are consistent with the effective tube-end opening side-effect of the SWCNT purification
\cite{HiraharaPRB}\cite{KatauraSM}. Additional heat treatment in air or refluxing in H$_{2}$O$_{2}$ 
does not increase the fullerene filling levels in our samples. The mean value, $d$, and the variance, $\sigma$, 
of the tube diameters were determined from multifrequency Raman measurements\cite{KuzmanyEPJB} and it was found that $d$ = 1.50 nm, $\sigma$ = 0.1 nm and $d$ = 1.34 nm, 
$\sigma$ = 0.09 nm for the NCL- and purified Rice-SWCNT samples, respectively. 
We followed the method of Kataura et al. \cite{KatauraSM} to fill fullerenes from the vapor phase, 
denoted as vapor-filling in the following. This involves sealing of the SWCNT material with the 
fullerene in a quartz ampoule after degassing at 300 $^{o}$C and keeping it at 650 $^{o}$C for 2 hours. 
The resulting material was sonicated in toluene in order to remove non-reacted 
fullerenes, filtered, and dried from toluene at 400 $^{o}$C in dynamic vacuum. Dynamic 
vacuum treatment at 700 $^{o}$C is equivalent to this last step in removing non-reacted fullerene particles 
without an observable effect on the peapods. Fullerene filling into SWCNT in n-hexane, denoted as 
solvent-filling in the following, is achieved with mixing typically 5 mg of the SWCNT material 
with 10 ml n-hexane with 5 mg of C$_{60}$ or C$_{70}$. The as-received NCL-SWCNT materials were dried by the manufacturer 
and care was taken to keep it away from humidity. The 400 $^{o}$C dynamic vacuum degassing of the Rice-SWCNT 
was also crucial for the solvent-filling: rinsing it in water prevents any further 
solvent-fillability probably because water enters into the nanotubes. The SWCNT, fullerene and n-hexane mixture 
was sonicated for 5 minutes resulting in the partial dissolution of 
C$_{60}$ due to the relatively low room temperature solubility, 0.043 mg/ml\cite{Dresselhaus}, of C$_{60}$ 
in n-hexane. The C$_{60}$ solution, undissolved C$_{60}$ and SWCNT mixture was then refluxed at 69 $^{o}$C 
for 2 hours. After this treatment, the filtered bucky-papers were dried in air at 
120 $^{o}$C for 1 hour. Not encapsulated C$_{60}$ that covers the bucky-paper is removed with the two 
methods mentioned above: sonication in toluene or by dynamic vacuum treatment at 700 $^{o}$C. Our studies have shown 
that both methods yield identical materials. The same steps were followed for the production of the 
C$_{70}$@SWCNT peapod material. 
DWCNT transformation of the peapod samples was performed with a 2 h long dynamic vacuum treatment at 1250 $^{o}$C 
following Ref.\cite{BandowCPL}. \smallskip

\textit{Transmission electron microscopy.} High resolution transmission electron microscopic (HR-TEM) studies 
were performed on a TECNAI F20 field emission microscope equipped with a Gatan Image Filter (GIF 2001) operated 
at 120kV or 200 kV. Electron transparent samples were prepared by drying a suspension of peapod material and 
N,N-Dimethylformamide on a holey carbon grid. \smallskip

\textit{Raman spectroscopy.} Multi frequency Raman spectroscopy was studied on a Dilor xy triple spectrometer 
equipped with a cryostat for the 20-600 K temperature range in the 1.83-2.54 eV (676-488 nm) energy range using 
an Ar-Kr mixed-gas laser. We used Raman spectroscopy to characterize the diameter distribution of the 
SWCNT, to determine the peapod concentration, and to monitor the DWCNT transformation of the peapod samples.\smallskip

\textit{Electron energy loss spectroscopy.} EELS was performed in 
transmission at a 170 keV in a purpose built spectrometer which combines both high energy (180 meV) and 
momentum resolution (0.6 nm$^{-1}$). Details can be found in Ref. \cite{FinkAEEP} and references therein. 
Free standing films of about 100 nm effective thickness, which is thin enough to avoid multiple scattering, were 
prepared as described in Ref.\cite{LiuPRB}.\smallskip

\textit{X-ray studies.} X-ray diffraction images were measured with Cu K$_{\alpha}$ radiation from a rotating 
anode X-ray 
generator and a pinhole camera, equipped with a two-dimensional, position-sensitive detector \cite{PeterlikCar}.
 The system works in a 10$^{-4}$ mbar vacuum to avoid parasitic scattering from air. 
Radial averages of the two-dimensional spectra were evaluated to obtain the scattering curves 
$q=4 \pi / \lambda sin \theta$, with $2\theta$ being the scattering angle and $\lambda=0.1542$ nm being the 
X-ray wavelength. The strong increase in scattering intensity, that is always observed for SWCNT towards small
$q$, is subtracted by a power-law \cite{Xraysubstr}.\smallskip

\textit{Electron spin resonance spectroscopy.} The N@C$_{60}$:C$_{60}$ endohedral fullerene:fullerene solid 
solution was produced in a N$_{2}$ arc-discharge tube following Ref.\cite{PietzakCPL} with a typical yield of 
1-10 ppm \cite{JanossyKirch2000}. After the solvent filling steps, excess fullerenes were removed by sonication
in toluene and the filtered peapod material was dried at 100 $^{o}C$ in air. 
The peapod and the reference SWCNT materials were mixed with the ESR silent high purity SnO$_{2}$ 
in a mortar to separate the pieces of the conducting bucky-papers. The samples were sealed under dynamic vacuum. 
A typical microwave power of 10 $\mu$W and 0.01 mT magnetic field modulation at ambient 
temperature were used for the measurements in a Bruker Elexsys X-band spectrometer. \smallskip

\begin{figure}[tbp]
\includegraphics[width=0.8\hsize]{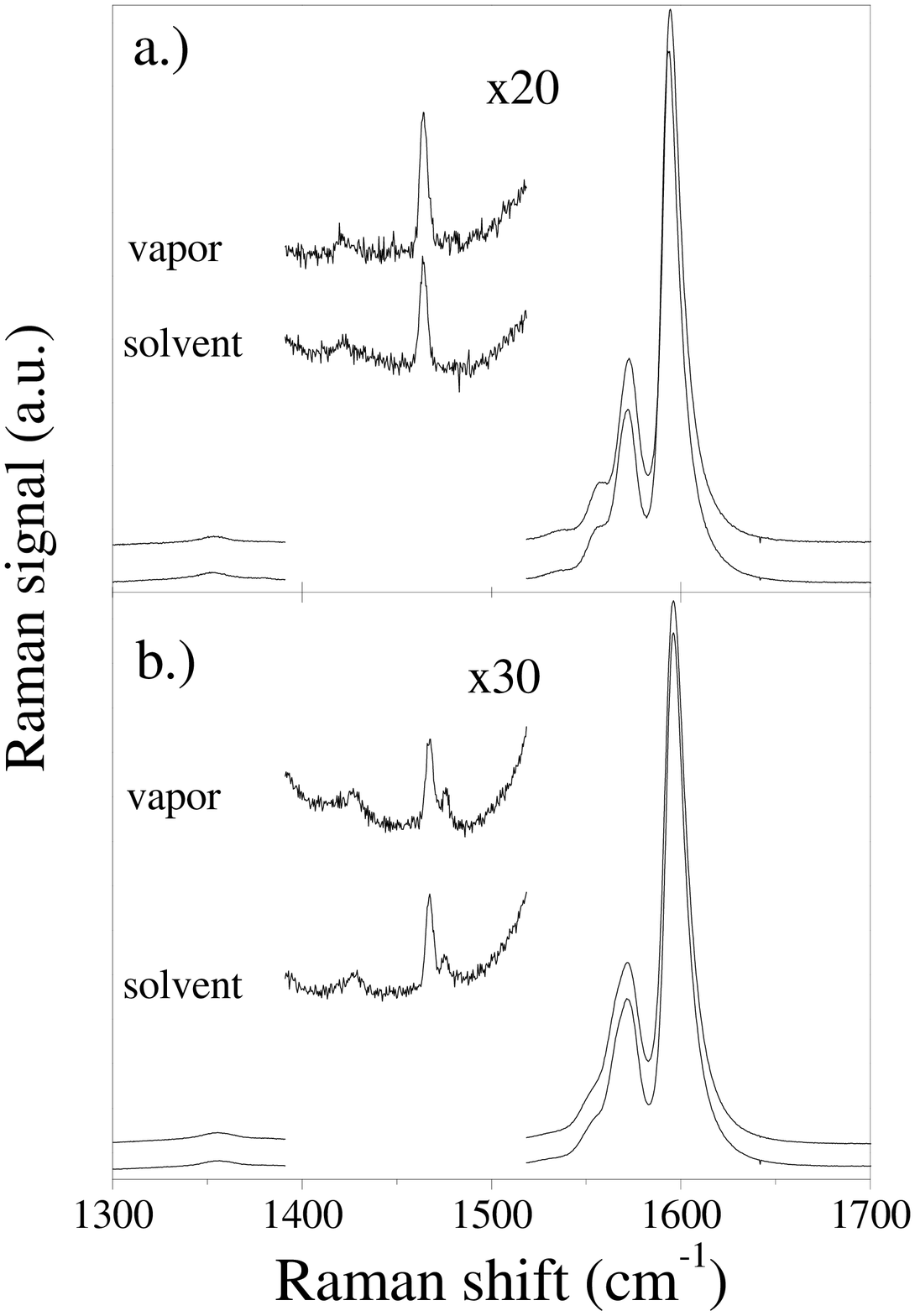}
\caption{a.) Raman spectra of vapor and solvent filled C$_{60}$@NCL-SWCNT and b.) vapor and solvent 
filled C$_{60}$@Rice-SWCNT at $\lambda$ = 488 nm and 90 K. The spectra are normalized to the amplitude of the 
SWCNT G mode.}
\label{peapodspectra}
\end{figure}

\section{Results and discussion}
We performed TEM studies on our vapor and solvent prepared samples. For both kinds of materials TEM micrographs 
(not shown) proved that an abundant peapod concentration was achieved. However, it is not representative of the 
bulk of the sample thus filling efficiency has to be determined from spectroscopic investigations. In Figure 1., 
we show the comparison of the Raman spectra of vapor- and solvent-filled C$_{60}$ peapod samples. The Raman 
spectra of peapod samples in the plotted frequency range consist of the SWCNT G and D modes 
at 1590 and 1355 cm$^{-1}$, respectively and narrow lines related to the Raman active modes of the C$_{60}$ inside 
the SWCNT\cite{KatauraSM}\cite{PichlerPRL}. We show enlarged the most significant Raman active line of C$_{60}$ 
peapod, the A$_{g}$(2) mode. For the NCL-SWCNT peapod sample, we observe a single A$_{g}(2)$ line at 1466 cm$^{-1}$ 
and for the Rice-SWCNT the well known doublet peapod signal at 1466 and 1474 cm$^{-1}$\cite{PichlerPRL}. 
The absence of an extra line at 1469 cm$^{-1}$, that is the Raman shift of the A$_{g}(2)$ of crystalline C$_{60}$ 
and the observation of the characteristic double A$_{g}(2)$ line structure in the Rice-SWCNT peapod sample 
are evidence that no C$_{60}$ is present apart from those encapsulated in the nanotubes. 
The larger mean tube diameter of the NCL-SWCNT is consistent with the absence of the weaker satellite of the 
A$_{g}(2)$ at 1474 cm$^{-1}$ in this sample as this signal is associated with the presence of immobile C$_{60}$ 
molecules in the smaller diameter nanotubes \cite{Pfeifferunpub}. 
We have also observed that the Raman spectra of vapor and solvent prepared C$_{70}$@SWCNT (spectra not shown) 
are identical also.

We determined the peapod concentration quantitatively from EELS measurements using the method described in 
Ref. \cite{LiuPRB} for both the vapor and solvent prepared samples. 
The C1s core level spectrum (not shown) contains contributions from 
carbon in C$_{60}$ and in the SWCNT starting material and additional carbon in the sample. When compared to a 
non C$_{60}$ encapsulating SWCNT reference, the excess C1s signal related to C$_{60}$ can be determined in the 
peapod samples. This yields the number of C$_{60}$ related carbon atoms that is directly translated to the 
filling level. Carbon in some impurity phases also affects the measurement. Thus, the values provided by this 
technique are only higher limits of the C$_{60}$ filling. Nevertheless, the current samples are of comparable 
quality than in the previous study\cite{LiuPRB} so any effect related to impurities gives the same order of 
error. Following Ref. \cite{LiuPRB}, the C$_{60}$ related C1s contributions can be translated to the volume 
filling with simple geometrical considerations taking into account the mean tube diameter and the 0.97 nm 
interfullerene distance\cite{HiraharaPRB}. We found that within experimental precision both the vapor and 
solvent prepared peapod materials have an overall C$_{60}$ filling of 55$\pm $5 \% for the NCL-SWCNT. 
This has to be compared with 61$\pm $5 \% found in a previous study on highly filled samples prepared with the 
vapor method \cite{LiuPRB}. The Rice-SWCNT has a smaller peapod content as seen from the Raman spectra that is 
most probably related to the smaller tube diameters in this sample as C$_{60}$ can only enter into nanotubes 
with $d$ $\gtrsim $ 1.2 nm\cite{LiuPRB}. Nevertheless, the similar peapod content of the vapor and solvent prepared 
materials emphasizes the effectiveness of the solvent filling method.

\begin{figure}[tbp]
\includegraphics[width=0.8\hsize]{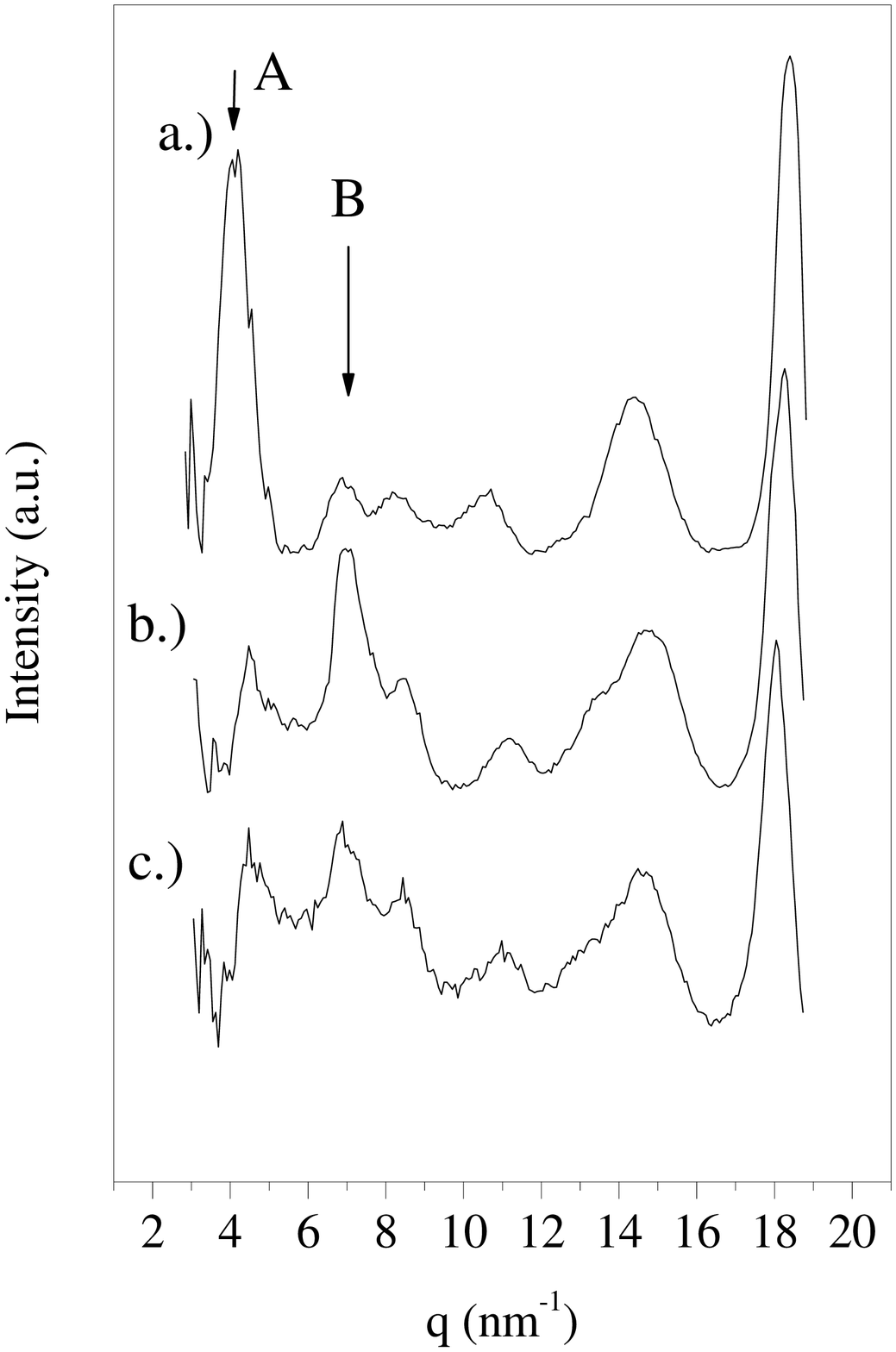}
\caption{X-ray diffraction profile of a.) pristine NCL-SWCNT, b.) vapor and c.) solvent prepared 
C$_{60}$@NCL-SWCNT samples. The relative intensity of the A and B peaks is a measure of the peapod concentration. 
The scattering peak observed at 18 nm$^{-1}$ comes from residual graphitic carbon.}
\label{xray}
\end{figure}

We also checked the consistency of the peapod filling content from X-ray diffraction. In Fig. 2., X-ray 
diffraction patterns of the pristine NCL-SWCNT, the peapod sample prepared by the vapor and the solvent 
method are shown. The encapsulation with C$_{60}$ strongly modifies the intensity of the peaks, 
in particular the 10 peak around $q = 4.5 $ nm$^{-1}$ is strongly depressed in comparison to the other peaks \cite{AbePRB}. 
The intensity modulation of the hexagonal lattice peaks by the different form factor of filled and empty SWCNT 
can be related to the filling content of the encapsulated peapods \cite{AbePRB}\cite{KatauraAPA}. 
The first peak (A) decreases for filled nanotubes in comparison to the pristine SWCNT material and a 
second peak (B) appears, which is related to the one-dimensional lattice of the encapsulated C$_{60}$ molecules. 
The X-ray spectra of the vapor and the solvent prepared peapod samples are nearly identical, which proves 
that a similar, high peapod filling is achieved with the two methods. Moreover, it proves that 
the solvent prepared peapod samples possess a similarly well ordered one-dimensional peapod structure.
A numerical simulation following 
\cite{AbePRB}\cite{KatauraAPA} leads for the inter-C$_{60}$ distance and the coherence length of 0.95 nm and 25 nm, respectively,
in agreement with previous electron diffraction studies \cite{LiuPRB}.

\begin{figure}[tbp]
\includegraphics[width=0.8\hsize]{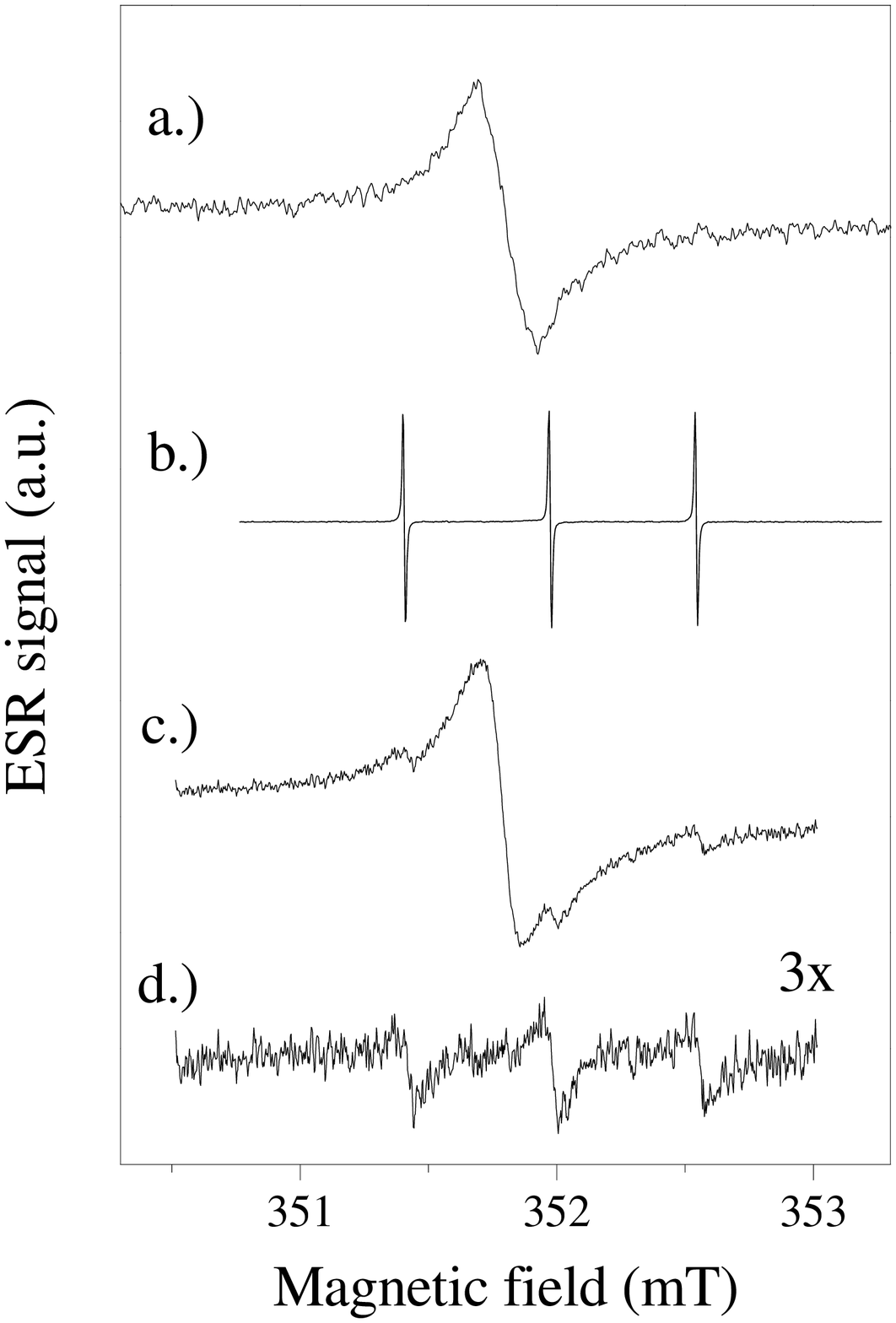}
\caption{X-band electron spin resonance spectrum of the a.) pristine SWCNT, b.) crystalline N@C$_{60}$:C$_{60}$, 
c.) (N@C$_{60}$:C$_{60}$)@SWCNT and d.) the triplet component of the (N@C$_{60}$:C$_{60}$)@SWCNT ESR spectrum 
at ambient temperature.}
\label{esr}
\end{figure}

In Fig. 3., we show the ESR spectra of the starting NCL-SWCNT, (N@C$_{60}$:C$_{60}$)@NCL-SWCNT, and 
N@C$_{60}$:C$_{60}$. The ESR spectrum of the pristine NCL-SWCNT for the magnetic field range shown is 
dominated by a signal that is assigned to some residual carbonaceous material, probably graphite. 
Fig. 3c. shows, that after the solvent encapsulation of N@C$_{60}$:C$_{60}$ in the NCL-SWCNT, we observe a 
hyperfine N triplet ESR, similar to that in pristine N@C$_{60}$:C$_{60}$, superimposed on the broad signal 
present in the pristine nanotube material. Fig. 3d. shows the triplet component of this signal after 
subtracting the signal observed in pristine SWCNT. The hiperfine triplet in N@C$_{60}$:C$_{60}$ is the result 
of the overlap of the $^{4}$S$_{3/2}$ state of the three 2p electrons of the N atom and the $^{14}$N nucleus, 
with nuclear spin, I =1. The isotropic hyperfine coupling of N@C$_{60}$:C$_{60}$ is unusually high as a result 
of the strongly compressed N atomic 2p$^{3}$ orbitals in the C$_{60}$ cage thus it unambiguously identifies 
this material\cite{WeidingerPRL}. The hyperfine coupling constant observed for the triplet structure in the 
encapsulated material, $A$ =0.57$\pm $0.01 mT, agrees within experimental precision with that observed in 
N@C$_{60}$:C$_{60}$\cite{WeidingerPRL}, which proves that the encapsulated material is 
(N@C$_{60}$:C$_{60}$)@SWCNT. The preparation procedure, as detailed above, guarantees that non-encapsulated 
N@C$_{60}$ is not present in our samples. The ESR linewidth for the encapsulated material, 
$\Delta H_{pp}$ = 0.07 mT, is significantly larger than the resolution limited $\Delta H_{pp}$ =0.01 mT in the 
pristine N@C$_{60}$:C$_{60}$ material, the lines being Lorentzian. The most probable cause for the broadening 
is static magnetic fields from residual magnetic impurities in the SWCNT\cite{TangSCI}. 
The ESR signal intensity is proportional to the number of N spins, and this allows the quantitative comparison 
of N concentrations in (N@C$_{60}$:C$_{60}$)@SWCNT and N@C$_{60}$:C$_{60}$. Taking the filling value from the 
EELS measurement, we obtain that the N spin concentration in (N@C$_{60}$:C$_{60}$)@SWCNT is $\sim $2.5 times 
smaller than in the starting N@C$_{60}$:C$_{60}$ material. This cannot be due to a loss of N spins during 
the synthesis as N@C$_{60}$:C$_{60}$ is stable below 100 $^{o}$C and N loss becomes 
rapid above 220 $^{o}$C only\cite{WaiblingerPRB}. However, the observed difference may be caused by the limited 
accuracy of the ESR intensity measurement also caused by microwave losses as the conducting bucky-paper pieces 
are separated from each-other in a poorly controlled way.

\begin{figure}[tbp]
\includegraphics[width=0.8\hsize]{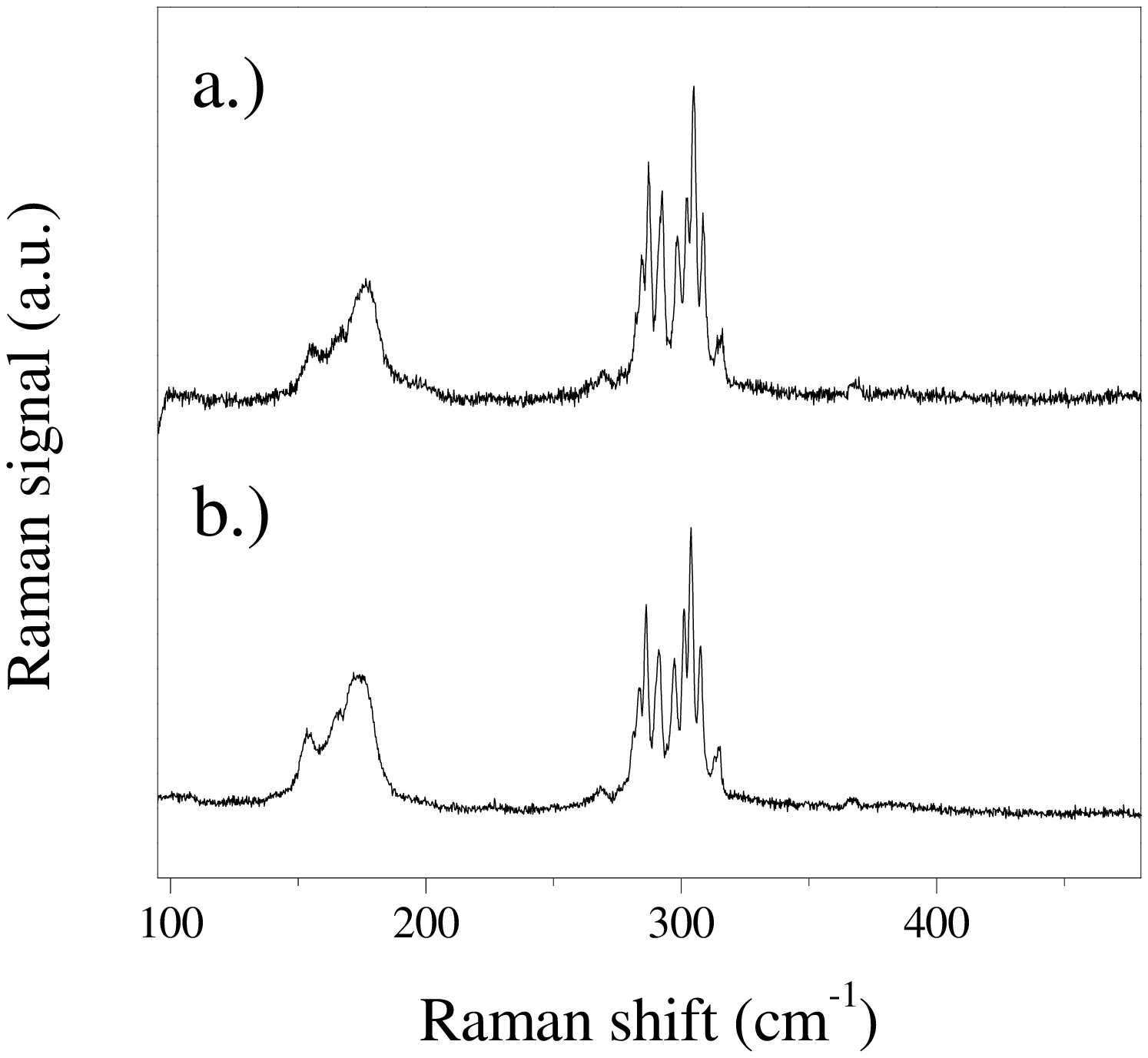}
\caption{Raman spectra of DWCNT obtained from a.) vapor and b.) solvent filled C$_{60}$@NCL-SWCNT at 
$\lambda$ = 676 nm and ambient temperature.}
\label{dwcnt}
\end{figure}

In Figure 4., we show the Raman spectra of DWCNT prepared from vapor and solvent prepared C$_{60}$@NCL-SWCNT 
peapod samples. For both spectra, the narrow Raman lines in the 250-350 cm$^{-1}$ spectral range correspond 
to the inner tubes\cite{PfeifferPRL}. Similar patterns were observed at other laser lines (not shown) proving 
that the inner tube diameter distributions are identical in the two kinds of samples irrespective of the 
peapod synthesis method. This shows that peapod materials can be produced with the solvent method that are 
suitable for the production of DWCNT materials. 

\section{Conclusion}
We presented the preparation of fullerene encapsulated SWCNT at temperatures slightly above room temperature. 
This method produces very high filling and provides a simple alternative to the commonly used vapor filling method. 
Its advantage is the relative simplicity and the possibility to scale it up to larger amounts. 
We have encapsulated a temperature senstive 
material, N@C$_{60}$:C$_{60}$, and observed its ESR signal. To our knowledge it is the first successful magnetic 
resonance experiment on an SWCNT encapsulated spin-probe. It has been speculated that such materials, when 
available in higher spin concentrations, may be fundamental elements of quantum-computering\cite{HarneitPSS}. 
The solvent prepared peapod samples are transformed to DWCNT with a yield identical to that from vapor prepared 
materials. This opens the way for the production of high purity and highly perfect DWCNT in industrial amounts.

\section{Acknowledgement}
The authors gratefully acknowledge useful discussions with Christian Kramberger and Rudolf Pfeiffer. 
This work was supported by the Austrian Science Funds (FWF) project Nr. P16315 and Nr. 14893, by the EU project 
NANOTEMP BIN2-2001-00580, by the Deutsche Forschungsgemeinschaft DFG project PI 440, 
by the Hochschuljubil\"{a}umsstiftung der Stadt Wien project H-175/2001 and by the 
Hungarian State Grant OTKA T043255. The Budapest ESR Laboratory is member of the SENTINEL European 
Infrastructural Network.

\textit{Note added.} After preparing this manuscript for publication we learned about a similar 
low temperature C$_{60}$ encapsulation method from Ref.\cite{YudasakaCPL} using ethanol as solvent.


\begin{thebibliography}{9}
\bibitem{IijimaNAT} S. Iijima, Nature (London) 354 (1991) 56.

\bibitem{SmithNAT} B. W. Smith, M. Monthioux, and D. E. Luzzi, Nature (London) 396 (1998) 323.

\bibitem{MonthiouxCAR} M. Monthioux, Carbon 40 (2002) 1809.

\bibitem{BandowCPL} S. Bandow, M. Takizaw, K. Hirahara, M. Yudasaka, S. Iijima, Chem. Phys. Lett. 337 (2001) 48.

\bibitem{nanocarblab} http://www.nanocarblab.com

\bibitem{HiraharaPRB} K. Hirahara, S. Bandow, K. Suenaga, H. Kato, T. Okazaki, T. Shinohara, and S. Iijima, Phys. Rev. B. 64 (2001) 115420.

\bibitem{KatauraSM} H. Kataura, Y. Maniwa, T Kodama, K Kikuchi, K. Hirahara, K. Suenaga, S. Iijima, S. Suzuki, Y. Achiba,W. Kratschmer, Synth. Met. 121 (2001) 1195.

\bibitem{KuzmanyEPJB} H. Kuzmany, W. Plank, M. Hulman, C. Kramberger, A. Gruneis, T. Pichler, H. Peterlik, H. Kataura, Y. Achiba, Eur. Phys. J. B 22 (2001) 307.

\bibitem{Dresselhaus} M. S. Dresselhaus, G. Dresselhaus, P. C. Ecklund: Science of Fullerenes and Carbon Nanotubes, Academic Press, 1996.

\bibitem{FinkAEEP} J. Fink, Adv. Electron. Electron Phys. 75 (1989) 121.

\bibitem{LiuPRB} X. Liu, T. Pichler, M. Knupfer, M. S. Golden, J. Fink, H. Kataura, Y. Achiba, K. Hirahara, S. Iijima, Phys. Rev. B 65 (2002) 045419.

\bibitem{PeterlikCar} H.Peterlik, P.Fratzl, and K.Kromp, Carbon 32 (1994) 939.

\bibitem{Xraysubstr} A. Rinzler, J. Liu,H. Dai, P. Nikolaev,C.B. Huffman,F.J. Rodriguez-Macias, P.J. Boul, A.H. Lu, D. Heymann, D.T. Colbert, R.S. Lee, J.E. Fischer, A.M. Rao, P.C. Eklund, R.E. Smalley, Appl. Phys. A 67 (1998) 29.

\bibitem{PietzakCPL} B. Pietzak, M. Waiblinger, T.A. Murphy, A. Weidinger, M. Hohne, E. Dietel, A. Hirsch, Chem. Phys. Lett. 279 (1997) 259.

\bibitem{JanossyKirch2000} A. J\'{a}nossy, S. Pekker, F. F\"{u}l\"{o}p, F. Simon, G. Oszl\'{a}nyi, in H. Kuzmany, J. Fink, M. Mehring, S. Roth (Eds.) electronic Properties of Novel Molecular Nanostructures, AIP Conference Proceedings, New York, 2000, p. 199.

\bibitem{PichlerPRL} T. Pichler, H. Kuzmany, H. Kataura, Y. Achiba, Phys. Rev. Lett. 87 (2001) 267401.

\bibitem{Pfeifferunpub} R. Pfeiffer \textit{et al.}, unpublished.

\bibitem{AbePRB} M. Abe, H. Kataura, H. Kira, T. Kodama, S. Suzuki, Y. Achiba, K. Kato, M. Takata, A. Fujiwara, K. Matsuda, Y. Maniwa, Phys. Rev. B 68 (2003) 041405.

\bibitem{KatauraAPA} H. Kataura, Y. Maniwa, M. Abe, A. Fujiwara, T. Kodama, K. Kikuchi, J. Imahori, Y. Misaki, S. Suzuki, Y. Achiba, Appl. Phys. A 349 (2002) 349.

\bibitem{WeidingerPRL} T. Almeida Murphy, Th. Pawlik, A. Weidinger, M. Höhne, R. Alcala, J.-M. Spaeth, Phys. Rev. Lett. 77 (1996) 1075.

\bibitem{TangSCI} X.-P. Tang, A. Kleinhammes, H. Shimoda, L. Fleming, K. Y. Bennoune, S. Sinha, C. Bower, O. Zhou, Y. Wu, Science 288 (2000) 492.

\bibitem{WaiblingerPRB} M. Waiblinger, K. Lips, W. Harneit, A. Weidinger, E. Dietel, A. Hirsch, Phys. Rev. B 64 (2001) 159901.

\bibitem{PfeifferPRL} R. Pfeiffer, H. Kuzmany, C. Kramberger, C. Schaman, T. Pichler, H. Kataura, Y. Achiba, J. K\"{u}rti, V. Z\'{o}lyomi, Phys. Rev. Lett. 90 (2003) 225501.

\bibitem{HarneitPSS} W. Harneit, C. Meyer, A. Weidinger, D. Suter, J. Twamley, Phys. St. Solidi B 233 (2002) 453.

\bibitem{YudasakaCPL} M. Yudasaka, K. Ajima, K. Suenaga, T. Ichihashi, A. Hashimoto, S. Iijima, Chem. Phys. Lett. 380 (2003) 42.

\end{thebibliography}
\end{document}